\documentclass[a4paper,10pt]{llncs}
\usepackage{graphicx}
\usepackage{fancyvrb}

\usepackage{algorithmic}

\begin{document}
\title{Augmenting Agent Platforms to Facilitate Conversation Reasoning}

\author{David Lillis \and Rem W. Collier}
\institute{School of Computer Science and Informatics\\
University College Dublin\\
\email{david.lillis,rem.collier@ucd.ie}}

\maketitle

\pagestyle{empty}
\thispagestyle{empty}

\bibliographystyle{splncs}

\begin{abstract}

Within Multi Agent Systems, communication by means of Agent Communication Languages (ACLs) has a key role to play in the co-operation, co-ordination and knowledge-sharing between agents. Despite this, complex reasoning about agent messaging, and specifically about conversations between agents, tends not to have widespread support amongst general-purpose agent programming languages.

ACRE (Agent Communication Reasoning Engine) aims to complement the existing logical reasoning capabilities of agent programming languages with the capability of reasoning about complex interaction protocols in order to facilitate conversations between agents. This paper outlines the aims of the ACRE project and gives details of the functioning of a prototype implementation within the Agent Factory multi agent framework.
\end{abstract}

\section{Introduction} \label{sec:introduction}

Communication is a vital part of a Multi Agent System (MAS). Agents make use of communication in order to aid mutual cooperation towards the achievement of their individual or shared objectives. The sharing of knowledge, objectives and ideas amongst agents is facilitated by the use of Agent Communication Languages (ACLs). The importance of ACLs is reflected by the widespread support for them in agent programming languages and toolkits, many of which have ACL support built-in as core features.

In many MASs, communication takes place by way of individual messages without formal links between them. An alternative approach is to group related messages into conversations: ``task-oriented, shared sequences of messages that they observe, in order to accomplish specific tasks, such as a negotiation or an auction''~\cite{Labrou2001}.

This paper presents the Agent Conversation Reasoning Engine (ACRE). The principal aim of the ACRE project is to integrate interaction protocols into the core of existing agent programming languages. This is done by augmenting their existing reasoning capabilities and support for inter-agent communication by adding the ability to track and reason about conversations. Currently at the stage of an initial prototype, ACRE has been integrated with several agent programming languages running as part of the Common Language Framework of the Agent Factory platform~\cite{Collier2003}. The longer-term goals of ACRE include its use within other mainstream agent frameworks and languages.

The principal aim of this paper is to outline the goals of the ACRE project and to discuss its integration into Agent Factory.

This paper is laid out as follows: Section~\ref{sec:related} outlines some related work on agent interaction. Section~\ref{sec:overview} then provides an overview of the aims and scope of the ACRE project. The model used to reason about conversations is presented in Section~\ref{sec:model}. ACRE protocols are defined in an XML format that is outlined in Section~\ref{sec:xml}, followed by an example of a conversation in execution in Section~\ref{sec:example}. Details of the integration of ACRE into the Agent Factory framework are given in Section~\ref{sec:af}. Finally, Section~\ref{sec:conclusions} outlines some conclusions along with ideas for future work.

\section{Related Work} \label{sec:related}

In the context of Agent Communication Languages, two standards have found widespread adoption. The first widely-adopted format for agent communication was the Knowledge Query and Manipulation Language (KQML)~\cite{Finin1994}. An alternative agent communication standard was later developed by the Foundation for Intelligent Physical Agents (FIPA). FIPA ACL utilises what it considers to be a minimal set of English verbs that are necessary for agent communication. These are used to define a set of performatives that can be used in ACL messages~\cite{Poslad2000}. These performatives, along with their associated semantics, are defined in~\cite{FIPA00037}.

Recognising that one-off messages are limited in their power to be used in more complex interactions, FIPA also defined a set of interaction protocols~\cite{FIPA2000}. These are designed to cover a set of common interactions such as one agent requesting information from another, an agent informing others of some event and auction protocols. 

Support for either KQML or FIPA ACL communication is frequently included as a core feature in many agent toolkits and frameworks, native support for interaction protocols is less common. The JADE toolkit provides specific implementations of a number of the FIPA interaction protocols~\cite{Bellifemine2007}. It also provides a Finite State Machine (FSM) behaviour to allow interaction protocols to be defined~\cite{Dinkloh2004}. Jason includes native support for communicative acts, but does not provide specific tools for the development of agent conversations using interaction protocols. This is left to the agent programmer~\cite[p. 130]{Bordini2007}. A similar level of support is present within the Agent Factory framework~\cite{Collier2005}.

There do exist a number of toolkits, however, that do include support for conversations. For example, the COOrdination Language (COOL) uses FSMs to represent conversations~\cite{Barbuceanu1995}. Here, a conversation is always in some state, with messages causing transitions between conversation states. Jackal~\cite{Cost1998} and KaOS~\cite{Bradshaw1997} are other examples of agent systems making use of FSMs to model communications amongst agents. Alternative representations of Interaction Protocols include Coloured Petri Nets~\cite{Cost1999} and Dooley Graphs~\cite{Parunak1996}.

\section{ACRE Overview} \label{sec:overview}
ACRE is aimed at providing a comprehensive system for modelling, managing and reasoning about complex interactions using protocols and conversations. Here, we distinguish between \textit{protocols} and \textit{conversations}. A protocol is defined as a set of rules that dictate the format and ordering of messages that should be passed between agents that are involved in prolonged communication (beyond the passing of a single message). A conversation is defined as a single instance of multiple agents following a protocol in order to engage in communication. It is possible for two agents to engage in multiple conversations that follow the same protocol.

Such an aim can only be realised effectively if a number of features are already available. These include:

\begin{itemize}
   \item \textbf{Protocol definitions understandable by agents:} Interaction protocols must be declared in a language that all agents must be able to understand and share. This also has the advantage that the protocol definition is separated from its implementation in the agent, thus providing a programmer with a greater understanding of the format the communication is expected to take. ACRE uses an XML representation of a finite state machine for this purpose. This representation is further discussed in Section~\ref{sec:xml}. The separation of protocol definitions from agent behaviours also facilitates the development of external tools to monitor communication between agents.
   \item \textbf{Shared ontologies:} A shared vocabulary is essential to agents understanding each other's communications. A shared ontology defines concepts about which agents need to be capable of reasoning.
   \item \textbf{Plan repository:} With the two above features in place, an agent may reason about the sequence of messages being exchanged, as well as the content of those messages. This reasoning will typically result in an agent deciding to perform some action as a consequence of receiving certain communications. In this case, it is useful to have available a shareable repository of plans that agents may perform so that new capabilities may be learned from others. Clearly, the use of shared plans will be dependent on the agent programming language(s) being used.
\end{itemize}

The presence of these features aid greatly in the realisation of ACRE's aims. The principal aims are as follows:

\begin{itemize}
\item \textbf{External Monitoring of Interaction Protocols:} At its simplest level, conversation matching and recognition of interaction protocols allows for a relatively simple tool operating externally to any of the agents. This can intercept and read messages at the middleware level and is suitable for an open MAS in which agents communicate via FIPA ACL. This is a useful tool for debugging purposes, allowing developers to monitor communication to ensure that agents are following protocols correctly. This is particularly important where conversation management has been implemented in an ad-hoc way, with incoming and outgoing messages being treated independently and without a strong notion of conversations. In this case, the protocol definitions can be formalised after the implementation of the agents without interfering with the agent code itself until errors are identified.

\item \textbf{Internal Conversation Reasoning:} On receipt of a FIPA ACL message, it should be possible for an agent to identify the protocol being followed by means of the \texttt{protocol} parameter defined in the message (for the specification of the parameters available in a FIPA ACL message see~\cite{FIPA2002a}). Similarly, the initiator of a conversation should also set the \texttt{conversation-id} parameter, which is a unique identifier for a conversation. By referring to the the protocol identifier, an agent can make decisions about its response by consulting the protocol specification. Similarly, the conversation identifier may be matched against the stored history of ongoing conversations.

ACRE provides an agent with access to information about the conversations to which it is a party. This allows the agent to reason about this according to the capabilities of the agent programming language being used. One example of this is the use of this information to analyse the status of conversations and generate appropriate goals for the agent to successfully continue the conversation along the appropriate lines for the protocol that is specified. This has previously been done with the AFAPL2 agent programming language~\cite{Lillis2010a}, following the use of goals in~\cite{Braubach2007}. Goals represent the motivations of the participants in a conversation. Thus the agents' engagement in a particular conversation is decoupled from the individual messages that are being exchanged, allowing greater flexibility in reasoning about their reactions and responses.

\item \textbf{Organisation of Incoming Messages:} It is possible that an agent communicating with agents in another system may receive messages that do not specify their protocol and/or conversation identifier. In this case, it is useful for the agent to have access to definitions of the protocols in which it is capable of engaging so as to match these with incoming messages so as to categorise the messages.

In this situation, message fields such as the sender, receiver, message content and performative can be compared against currently active conversations to ascertain if it matches the expectations of the next step of the underlying protocol.

\item \textbf{Agent Code Verification:} The ultimate aim of ACRE is to facilitate the verification of certain aspects of agent code. In particular, given integration of conversation reasoning into a programming language, it should be possible to verify whether or not an agent is capable of engaging in a conversation following a particular protocol.
\end{itemize}

\section{Conversation Management} \label{sec:model}

ACRE models protocols as FSMs, with the transitions between states triggered by the exchange of messages between agents participating in the conversation. Messages, protocols and conversations are represented by tuples. As a FSM, each protocol is made up of states and transitions, which are also represented by tuples. This section presents these representations and also provides an informal description of the conversation management algorithm used within ACRE.

A \textit{message} is represented by the tuple $(s,r,c,\phi,p,x)$, where $s$ is the agent identifier of the message's sender, $r$ is the agent identifier of the recipient, $c$ is the conversation identifier, $\phi$ identifies the protocol, $p$ is the message performative and $x$ is the message content.

Each \textit{protocol} is represented by a tuple $(\phi, S, T, i, F)$ where $\phi$ is the protocol's unique identifier, $S$ and $T$ are sets of states and transitions respectively, $i$ is the name of the initial state and $F$ is a set of names of final (terminal) states.

Within these conversations, each \textit{state} is represented by the tuple $(n,s,\phi)$ where $n$ is the name of the state, $s$ is the status of the state (whether it is a start, end or intermediate state) and $\phi$ is the identifier of the protocol it belongs to. A \textit{transition} is represented by $(\sigma,\epsilon,s,r,p,x)$. Here, $\sigma$ and $\epsilon$ are the names of the start and end states respectively, $s$ and $r$ are the agent identifiers of the sending and receiving agents respectively, $p$ is the performative of the message triggering the transition and $x$ is the message content.

\begin{figure}[!htb]
\includegraphics[width=340px]{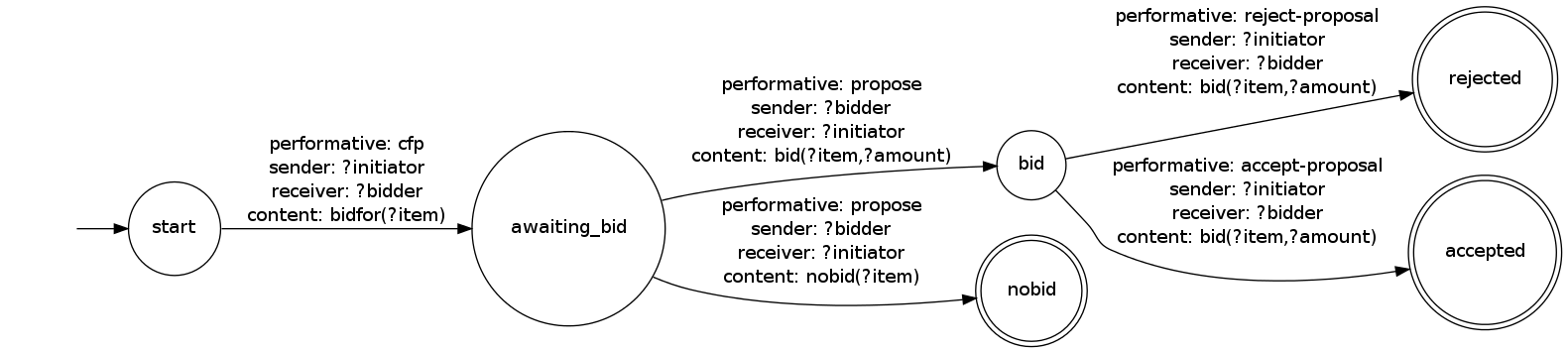}
\caption{FSM representation of the Vickrey Auction protocol}
\label{fig:fsm-vickrey}
\end{figure}

As a FSM, a protocol can easily visualised as shown in Figure~\ref{fig:fsm-vickrey}. This figure shows a FSM for a simple, one-shot Vickrey-style auction. It shows the states and transitions associated with this protocol. Transitions are triggered by comparison with messages exchanged between the participating agents.

Finally, a \textit{conversation} may be represented by $(\phi,A,s,c,B,\psi)$ where $\phi$ is the protocol identifier, $A$ is the set of participating agents, $s$ is the name of the conversation's current state, $c$ is the conversation identifier, $B$ is the current set of variable/value bindings and $\psi$ is the conversation status (active, completed or failed).

The values permitted in the tuples shown here are based on first-order logic, meaning that all values are constants, variables or functions. When considering whether a message is capable of advancing a conversation, its fields are compared with the corresponding elements of the conversation's available transitions.

When comparing values, the following rules apply: 
\begin{itemize}
\item Constant values match against other identical constant values (e.g. in Figure~\ref{fig:fsm-vickrey}, the first transition can only be triggered by a message with the performative \texttt{cfp}).
\item Variables match against any value.
\item Functions match other functions that have the same functor, have the same number of arguments and whose arguments in turn match.
\end{itemize}

In the pseudocode that follows in Figures~\ref{fig:candidate},~\ref{fig:candidate_protocol} and~\ref{fig:advance}, this is encapsulated by the function \texttt{matches(a,b)}.

The bindings associated with the conversation ($B$) is a set of key/value pairs that binds variables to constants or functions against which that they have been matched in triggering a transition. Any variables that have been matched against a constant or function in a triggering message are given a binding that is stored in $B$. In the example from Figure~\ref{fig:fsm-vickrey}, the sender of the initial message will have their agent identifier bound to the \texttt{?initiator} variable, so any further messages must be sent by/to that same agent, whenever the \texttt{?initiator} variable is used. This is an example of a variable being used in \textit{immutable context}. Once the variable has been bound to a value, that value may not change for the duration of the conversation.

An alternative approach is to use a variable in a \textit{mutable context}. In this situation, a variable may acquire a binding to a new value regardless of whether it has been previously bound. Further explanation (and examples) of the different variable contexts is presented in Section~\ref{sec:variables}. One special-case variable also exists. The \textit{anonymous variable} (denoted by ``\texttt{?}'') may not acquire any binding. Thus it acts as a wildcard match that will match against any values.

The following sections outline the three key stages of the conversation management algorithm. By convention, elements of tuples are denoted by using subscripts (e.g. the initial state ($i$) of a protocol ($p$) is shown as $p_i$).

\subsection{Identifying Candidate Conversations}

The first stage of the conversation management algorithm is carried out whenever a message is exchanged and is shown in Figure~\ref{fig:candidate}. This identifies any active conversations that may be advanced by a message that has been exchanged. If the message contains a defined conversation identifier (which are unique), then only a conversation bearing that identifier may be advanced by the message. In the event that a message is exchanged without a conversation identifier being present, any conversation with an available transition that may be triggered by the message will be considered a candidate.

A conversation can be advanced by a message if the elements of the message match against the corresponding elements of any available transitions (i.e. that begin at the current state of the conversation). If the message contains a defined conversation identifier, but that conversation cannot be advanced by the message, the status of the conversation must be changed to \textit{failed}.

\begin{figure}
\begin{algorithmic}
\STATE $C \gets \emptyset$ to store candidate conversations
\STATE $m \gets$ message sent/received
\FOR{\textbf{each} active conversation ($c$)}
   \IF{$m_c = c_c$ \OR $m_c = \bot$}
     \FOR{\textbf{each} transition ($t$) \textbf{where} $t_\sigma = c_s$}
        \IF{$matches(m_s, apply(c_B,t_s))$ \AND $matches(m_r,apply(c_B,t_r))$ \\ \hspace{0.5cm}\AND $matches(m_x,apply(c_B,t_x))$ \AND $matches(m_p,t_p)$}
           \STATE Add $c$ to $C$
        \ENDIF
      \ENDFOR
   \ENDIF
   \IF{$m_c = c_c$ \AND $c \notin C$ }
      \STATE $c_\psi \gets failed$
   \ENDIF
\ENDFOR
\end{algorithmic}
\caption{Identifying candidate conversations}
\label{fig:candidate}
\end{figure}

The \texttt{apply(B,a)} function is used to apply a set of bindings (\texttt{B}) to a term (\texttt{a}). If \texttt{a} is a variable used in an immutable context for which a binding exists in \texttt{B}, then the bound value is returned. Otherwise, \texttt{a} is returned unaltered.

\subsection{Identifying Candidates for New Conversations}

If no active conversations may be advanced by the given message, the second stage is to identify whether the message is capable of initiating a conversation using a known protocol. This procedure is shown in Figure~\ref{fig:candidate_protocol}.

\begin{figure}[!hb]
\begin{algorithmic}
\IF{$|C|$ = 0}
  \FOR{\textbf{each} protocol ($p$)}
     \IF{$m_\phi=p_\phi$ \OR $m_\phi = \bot$}
        \FOR{\textbf{each} transition ($t$) where $t_\sigma = p_i$}
           \IF{$matches(m_s, t_s)$ \AND $matches(m_r,t_r)$ \AND $matches(m_x,t_x)$}
              \IF{$m_c = \bot$}
                 \STATE Add $(p_\phi, \{m_s,m_r\}, p_i, nextid(), \emptyset, active)$ to $C$
              \ELSE
                 \STATE Add $(p_\phi, \{m_s,m_r\}, p_i, m_c, \emptyset, active)$ to $C$
              \ENDIF
           \ENDIF
        \ENDFOR
     \ENDIF
  \ENDFOR
\ENDIF
\end{algorithmic}
\caption{Identifying candidate protocols for new conversations}
\label{fig:candidate_protocol}
\end{figure}
If the message contains a protocol identifier, then only the protocol with that identifier is considered. Otherwise, the message is compared against the initial transition of each available protocol. On finding a suitable protocol, a new conversation is created and added to the set of candidate conversations (\texttt{C}). If the message contained a conversation identifier, this is used as the identifier for the new conversation. Otherwise, a new unique conversation identifier is generated (by means of the \texttt{nextid()} function).

\subsection{Advancing the Conversation}

Having identified conversations that match against the given message, the system must advance a conversation, as appropriate. This is shown in Figure~\ref{fig:advance}. At this stage, events are raised to the agent layer to inform the agent of the outcome of the process. If the message was not capable of advancing or initiating any conversation, an ``unmatched'' event is raised. If there were multiple candidate conversations (which cannot be the case if conversation identifiers are defined for all messages), an ``ambiguous'' event is raised.

If one candidate conversation was identified, this is advanced to the next appropriate state. Its bindings must be updated (using the \texttt{getBindings(m,t)} function) to include bindings for variables in the transition that were matched against values in the message. The anonymous variable may not acquire a binding. This function does not discriminate between variables based on the context in which they are used. Both mutable and unbound immutable variables are free to acquire new bindings. If an immutable context variable has previously been bound to a value, it is this value that is used in matching the message to the transition (by means of the \texttt{apply(B,a)} function shown in Figure~\ref{fig:candidate}). As the message content is frequently a function of first order logic, any variables within that function that match against corresponding parts of the message content will also acquire bindings in the same way as standalone variables.

\begin{figure}[!h]
\begin{algorithmic}
\IF{$|C| = 1$}
   \STATE $c \gets$ the matched conversation in $C$
   \STATE $t \gets$ the transition matched by the message $m$
   \STATE $c_s \gets t_\epsilon$
   \STATE $c_B \gets c_B \cup getBindings(m,t)$
      \IF{$c_s$ is an end state}
         \STATE $c_\psi \gets completed$
         \STATE $raiseEvent(completed,c)$
      \ELSE
          \STATE $raiseEvent(advanced,c)$
      \ENDIF
   \ELSIF{$|C|=0$}
      \STATE $raiseEvent(unmatched,m)$
   \ELSE
      \STATE $raiseEvent(ambiguous,m)$
   \ENDIF
\end{algorithmic}
\caption{Advancing the conversation}
\label{fig:advance}
\end{figure}

\section{The ACRE XML Format} \label{sec:xml}

In ACRE, interaction protocols are modelled using an XML file that follows the ACRE XML protocol schema definition~\footnote{http://acre.lill.is/protocol.xsd}. A sample of an XML representation of a Vickrey Auction Interaction Protocol is given in Figure~\ref{fig:xml} (this is the same protocol as the FSM in Figure~\ref{fig:fsm-vickrey}). Each protocol is identified by a name, a name and a version number (contained in the \texttt{<namespace>}, \texttt{<name>} and \texttt{<version>} tags respectively). The version number is intended to prevent multiple agents attempting to communicate using different protocol implementations (e.g. if an error is discovered in an earlier attempt at modelling a particular protocol). The use of a namespace helps to avoid conflicts whereby various developers implement different models of similar protocols using the same name.

Each protocol is represented by a number of states and transitions, defined using \texttt{<state>} and \texttt{<transition>} tags respectively. Each state has only a ``name'' attribute, so that it can be referred to in the transitions. The type of state each represents (i.e. terminal, initial or other) can be found on the fly when the protocol is loaded. A state at which no transition ends is considered a start state. States at which no transitions begin are terminal states. The reason these are not expressly marked in the protocol definition is because of the ability to import other protocols, which is discussed in Section~\ref{sec:import}.

Transitions are more complex, as these are required to match messages so as to trigger a change in the state of a conversation. Each \texttt{<transition>} tag contains up to six attributes, many of which attempt to match against one field of a FIPA message. The attributes in this file correspond with the values in the tuple representing a transition in Section~\ref{sec:model}. In addition to these message fields, the ACRE conversation manager will also examine the \texttt{protocol-id} and \texttt{conversation-id} fields to match messages to particular conversations.

The attributes allowable in a \texttt{<transition>} tag are as follows:
\begin{itemize}
   \item \textbf{Performative:} This is a mandatory field that specifies the performative that a message must have in order to trigger this transition. The attribute value must be exactly equal to the performative contained in the message for this transition to be triggered. Variables are not permitted in this field.
   \item \textbf{From State:} Another mandatory field, this indicates the state from which this transition may be triggered. If the conversation is in another state then this rule cannot match. In the majority of cases, the attribute value must be the same as the name of a state that is contained either in the protocol itself or in an imported protocol.
   
   In addition to exact state names, regular expression matching is also permitted. If a regular expression is provided (indicated by beginning and ending the value with a forward slash), then this transition will be triggerable from any state that matches this regular expression. In practice, the protocol interpreter will duplicate this transition for each state name that matches, thus fitting with the model outlined in Section~\ref{sec:model}.

   \item \textbf{To State:} This is another mandatory field and is used to indicate the state that the conversation will be in upon successful triggering of this transition. As with the ``From State'' attribute, this should match the name of a state that is either part of the protocol or is imported. However, it may not contain a regular expression, as the conversation state after the sending of a message must be clearly defined.
   \item \textbf{Sender:} This indicates which agent should be the sender of the message. Although it is allowable to use a constant value for this attribute, this is unusual as it specifically restricts the protocol to an agent with a a particular identifier. Generally, this will use a variable to refer to particular agents. The same variable may be used throughout the protocol to indicate that particular messages should be sent by the same agent, as it will have acquired a bound value the first time it matches against an agent identifier.
   \item \textbf{Receiver:} This attribute functions in a similar way to ``Sender'', with the exception that it is the recipient of the message that is being matched.
   \item \textbf{Content:} This attribute relates to the actual content of the message. It may be a constant, a variable or a function that possibly combines the two. Figure~\ref{fig:xml} illustrates the use of a function in the \texttt{content} field in each of the transitions.
\end{itemize}

The ``Sender'', ``Receiver'' and ``Content'' attributes are optional in a protocol definition. In each case, the default value if one is not supplied is the anonymous ``\texttt{?}'' variable that matches any value.

\subsection{Importing Protocols} \label{sec:import}

One other feature of the ACRE XML format is the ability to import from other protocols. When this occurs, all of the states and transitions from the imported protocol are added to those of the protocol containing the \texttt{<import>} tag. This means that transitions in a protocol may refer to states that are not in the protocol itself but rather are in the imported protocol.

One example of a use for this is the ``Cancel'' meta-protocol that is included in all of the standard FIPA Interaction Protocols. This protocol always works in an identical way, regardless of what the main protocol being followed is: at any non-terminal stage of the conversation, the initiator of the original conversation may terminate the interaction by means of a \texttt{cancel} message. This meta-protocol can be extracted into a separate ACRE protocol that is imported by all other protocols that support it.

\subsection{Variable Bindings} \label{sec:variables}

As mentioned in Section~\ref{sec:model}, the definition of protocols in ACRE allows the use of three types of variable. The \textit{anonymous variable} ``\texttt{?}'' is an unnamed variable that is capable of matching against any value. As such, it can be considered to be a wildcard match. A transition whose \texttt{content} attribute is set to ``\texttt{?}'' can be triggered by a message with any content (assuming the other fields in the message match the specified transition).

\begin{figure}[htbp]
\begin{Verbatim}[frame=single]
<?xml version="1.0"?>

<protocol xmlns="http://acre.lill.is"
xmlns:xsi="http://www.w3.org/2001/XMLSchema-instance"
xsi:schemaLocation="http://acre.lill.is http://acre.lill.is/proto.xsd">

   <namespace>is.lill.acre</name>
   <name>acre-vickreyauction</name>
   <version>0.1</version>

   <states>
      <state name="start"/>
      <state name="awaiting_bid" />
      <state name="bid" />
      <state name="nobid"/>
      <state name="accepted"/>
      <state name="rejected"/>
   </states>

   <transitions>
      <transition performative="cfp"
                  from-state="start"
                  to-state="awaiting_bid"
                  sender="?initiator"
                  receiver="?bidder"
                  content="bidfor(?item)" />
      <transition performative="propose"
                  from-state="awaiting_bid"
                  to-state="bid"
                  sender="?bidder"
                  receiver="?initiator"
                  content="bid(?item,?amount)" />
      <transition performative="propose"
                  from-state="awaiting_bid"
                  to-state="nobid"
                  sender="?bidder"
                  receiver="?initiator"
                  content="nobid(?item)" />
      <transition performative="accept-proposal"
                  from-state="bid"
                  to-state="accepted"
                  sender="?initiator"
                  receiver="?bidder"
                  content="bid(?item,?amount)" />
      <transition performative="reject-proposal"
                  from-state="bid"
                  to-state="accepted"
                  sender="?initiator"
                  receiver="?bidder"
                  content="bid(?item,?amount)" />
   </transitions>
</protocol>
\end{Verbatim}
\caption{ACRE XML Representation of the Vickrey Auction Protocol}
\label{fig:xml}
\end{figure}

Two types of named variable are permitted: \textit{immutable named variables}, which have ''\texttt{?}'' as a prefix followed by the variable name (e.g. \texttt{?item}) and \textit{mutable named variables} that are prefixed with ``\texttt{??}'' (e.g. \texttt{??amount}).

Each named variable is in scope for the duration of a conversation and is associated with values as it is matched against the actual fields in messages that trigger transitions. Whenever a named variable is used in an immutable context, it may match any content if has not already acquired a value. However, once it has been matched to a value, it may only match that value for the duration of the conversation. For example, Figure~\ref{fig:xml} uses \texttt{?initiator} to denote the agent that begins the Vickrey Auction with the sending of the initial call for proposals. Initially this variable does not have a value associated with it, so it may match the name of the relevant agent (e.g. ``agent1''). However, once this has been done, the variable \texttt{?initiator} may only match the value ``agent1'' for the remainder of the conversation.

In some situations it is desirable to have variables whose values may change as the conversation progresses. For that reason, mutable named variables are also facilitated. The difference between a mutable and immutable named variable is that a mutable named variable can match against any content, regardless of whether it previously has a value associated with it. When this occurs, the existing value is overwritten with the new value that has been matched.

The motivation behind the use of mutable variables can be seen by examining the Vickrey Auction protocol shown above. This is a one-shot auction, so immutable variables are sufficient. However, implementing an iterated auction is made far more complex if all variables are immutable. The second transition shown in Figure~\ref{fig:xml} would be unsuitable for this, as the \texttt{?amount} variable is restricted to match only whatever the first value it is matched against. By changing the content field of this transition to \texttt{bid(?item,??amount)}, the variable relating to the amount is used in mutable context and so its value may change (although the item being bid for must remain the same). Thus each time this transition is triggered, the \texttt{amount} variable acquires a new value: namely the amount of the latest bid that was submitted. This can later be referred to using texttt{?amount} by the other agent that wishes to accept or reject the bid. A similar usage can be seen in the example presented in Section~\ref{sec:example}.

\section{ACRE Conversation Example}
\label{sec:example}

\begin{figure}[!htb]
\centering
\includegraphics[scale=0.4]{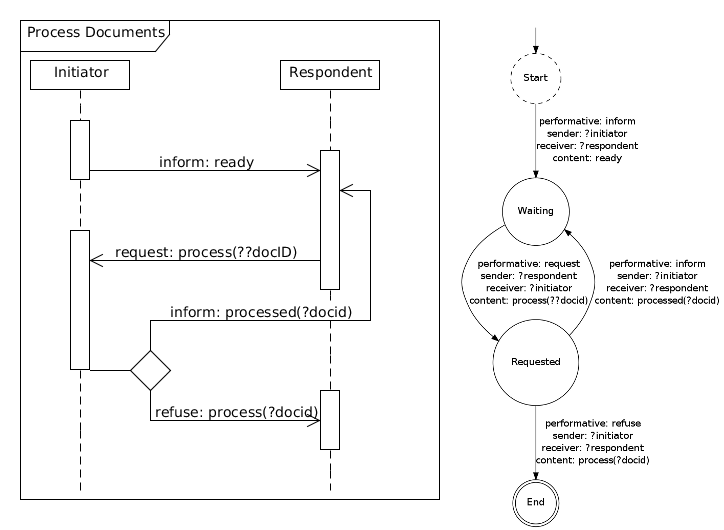}
\caption{Process Documents Protocol}
\label{fig:processdocuments-auml}
\end{figure}

This section presents an example of how an ACRE conversation may progress. The example is based on the ``Process Documents'' protocol shown in Figure~\ref{fig:processdocuments-auml}. This protocol is designed for a system where text documents must undergo some form of processing. The Initiator of the protocol is capable of performing this processing, though it must be made aware of which documents to work on by the Respondent.

Initially, the Initiator informs the Respondent that it is ready to process documents, to which the Respondent replies with a request to process a particular document. The Initiator may either process the document and inform the Respondent of this, or refuse to carry out this processing. In the former case, the Respondent will send the next document for processing and continue to do this until the Initiator eventually refuses.

Here, the Agent UML description shown to the left of Figure~\ref{fig:processdocuments-auml} is converted to the FSM shown on the right. The dashed line surrounding the ``Start'' state indicate that this is the current state initially.

The first transition contains constant values for both the performative and the content. This means that any message matching that transition must have those exact values in those fields. The message sender and receiver are variables, and since there are not yet any bindings associated with the conversation, these will match any agent identifiers in those fields in the message.

A message that will trigger this initial transition may look as follows (some unimportant fields have been omitted for clarity):

\begin{Verbatim}[samepage=true]
(inform
   :sender    processor
   :receiver  manager
   :content   ready
)
\end{Verbatim}

\begin{figure}[!htb]
   \begin{minipage}[b]{0.49\linewidth}
      \centering
      \includegraphics[scale=0.3]{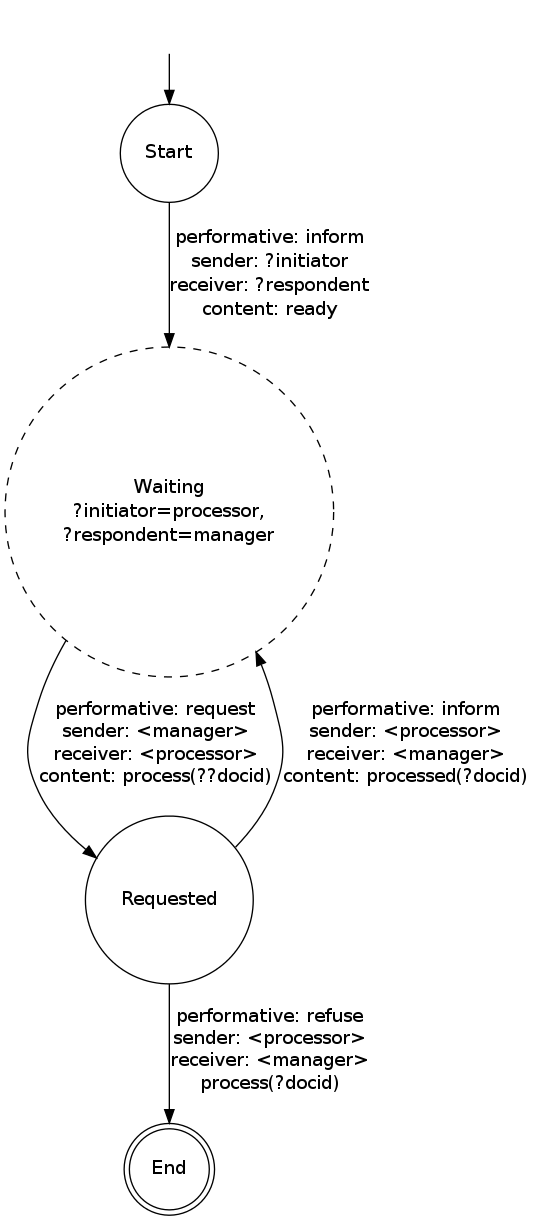}
      \caption{Process Documents protocol in the ``Waiting'' state}
      \label{fig:processdocuments-waiting1}
   \end{minipage}
   \hspace{0.05cm}
   \begin{minipage}[b]{0.49\linewidth}
      \includegraphics[scale=0.29]{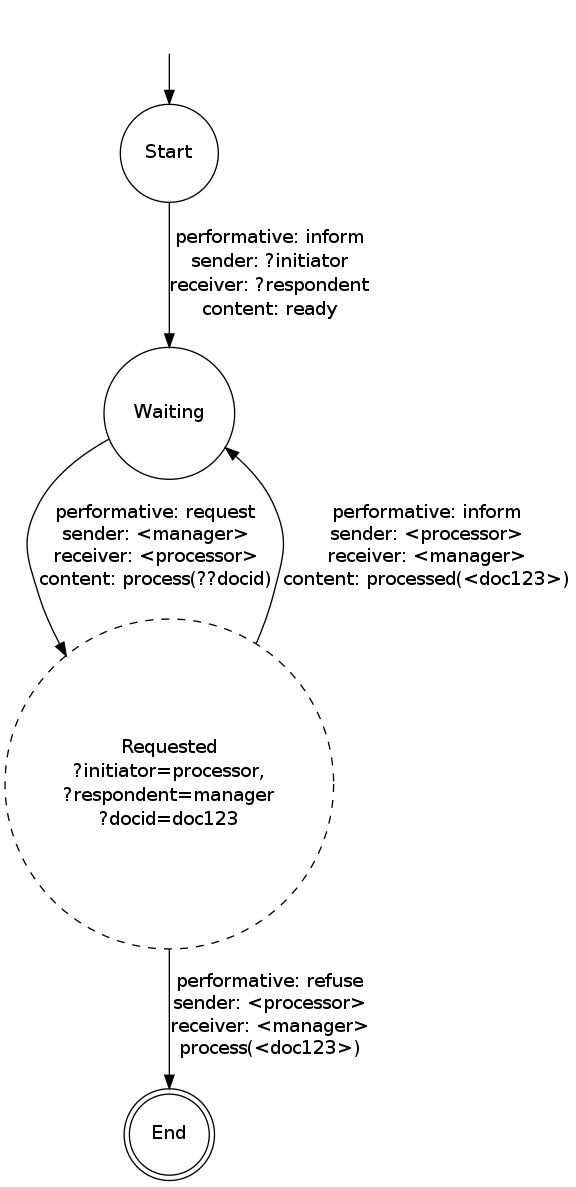}
      \caption{Process Documents protocol in the ``Requested'' state}
      \label{fig:processdocuments-requested1}
   \end{minipage}
\end{figure}

This results in the state of the conversation changing to ``Waiting'', as shown in Figure~\ref{fig:processdocuments-waiting1}. Because the \texttt{?initiator} and \texttt{?respondent} matched against the constants ``processor'' and ``manager'' respectively in the message, these bindings are associated with the conversation. As these variables are used in immutable context throughout, they must match their exact bound values for the remainder of the conversation. This is indicated in Figure~\ref{fig:processdocuments-waiting1} by replacing the variables with these values. Angle brackets have been placed around each of the replacement values in order to emphasise this.

The transition from the ``Waiting'' state can now only be triggered by a message sent by the manager agent to the processor agent with the ``request'' performative. The content must also match the transition, though with the use of the \texttt{??docid} variable, there is some flexibility in the values that can be matched.

In the next stage, the manager asks the processor to process the document with the identifier ``doc123''. This is done by means of the following message:

\begin{Verbatim}[samepage=true]
(request
   :sender    manager
   :receiver  processor
   :content   process(doc123)
)
\end{Verbatim}

As this message matches the transition, the conversation moves from the ``Waiting'' state to the ``Requested'' state, as shown in Figure~\ref{fig:processdocuments-requested1}.

\begin{figure}[!htb]
   \begin{minipage}[b]{0.49\linewidth}
      \centering
      \includegraphics[scale=0.29]{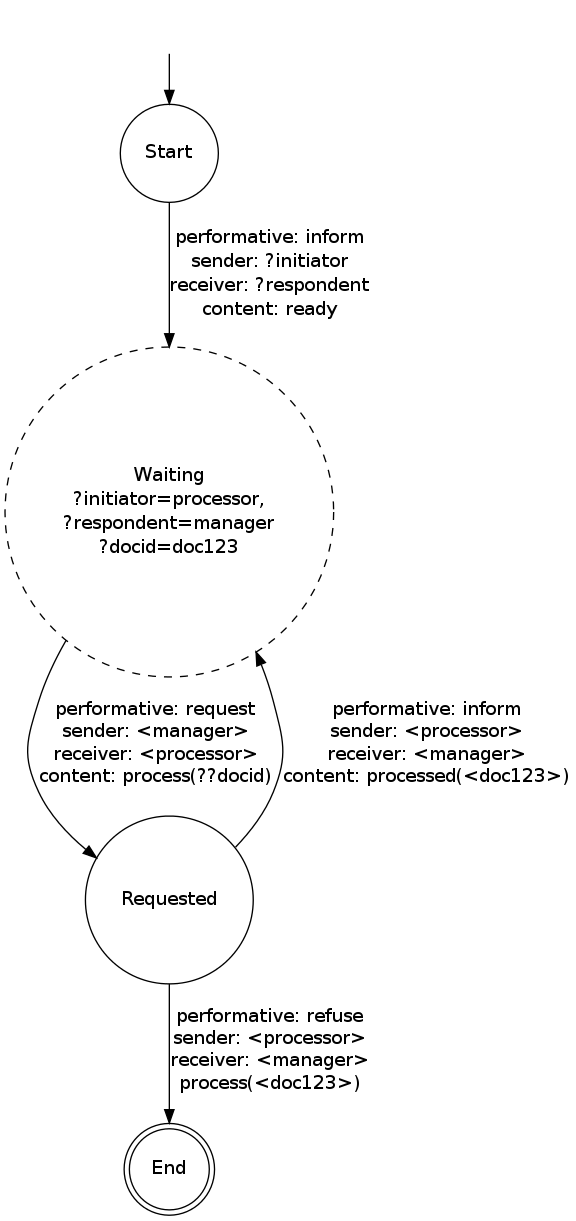}
      \caption{Process Documents protocol in the ``Waiting'' state for the second time.}
      \label{fig:processdocuments-waiting2}
   \end{minipage}
   \hspace{0.05cm}
   \begin{minipage}[b]{0.49\linewidth}
      \centering
      \includegraphics[scale=0.29]{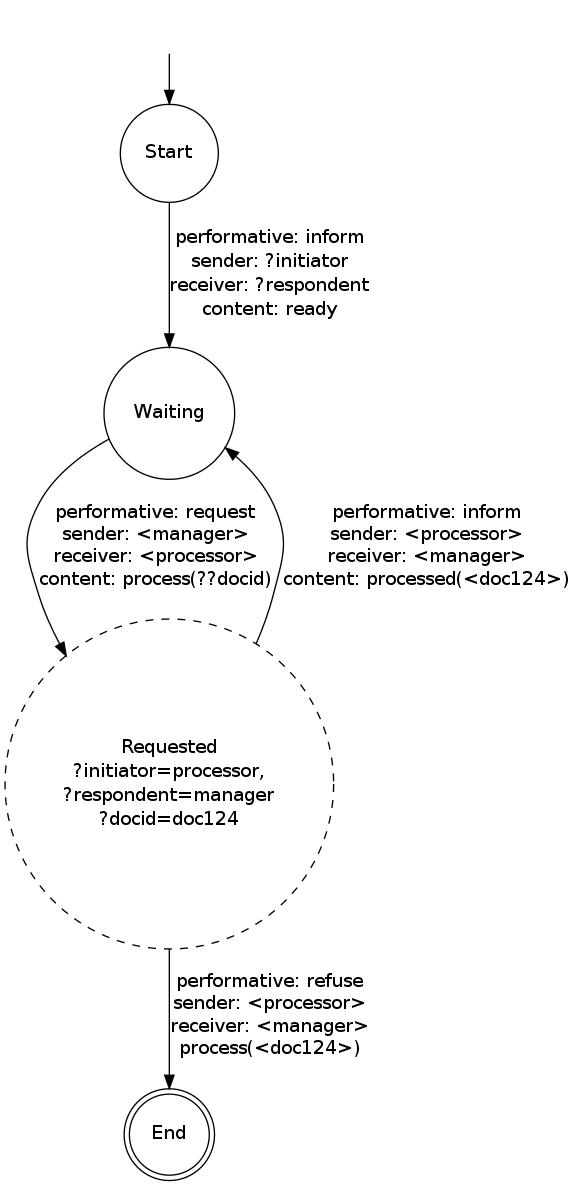}
      \caption{Process Documents protocol in the ``Requested'' state for the second time.}
      \label{fig:processdocuments-requested2}
   \end{minipage}
\end{figure}

At this point, the \texttt{?docid} variable has also acquired a binding, so this is replaced in all transitions using it in an immutable context. Again, this is indicated by the value being contained within angle brackets in Figure~\ref{fig:processdocuments-requested1}. At this point, the processor agent must either inform the manager that document ``doc123'' has been processed (thus returning the conversation to the ``Waiting'' state) or refusing to process that document. In each case, it is only the document identifier that has previously been bound to the \texttt{?docid} variable that may be used. Refusing to process a different document identifier would result in the conversation failing as the message could not match an available transition.

If the processor agrees to process the document and informs the manager of its completion, the conversation returns to the ``Waiting'' state, as shown in Figure~\ref{fig:processdocuments-waiting2}. Unlike the first time the conversation was in this state, this time the \texttt{?docid} variable has got a binding associated with it. However, the transition between ``Waiting'' and ``Requested'' uses this variable in a mutable context, meaning that it can match against any value contained in the next message. Thus the manager can ask the processor to process any document it wishes. In contrast, as noted previously, when moving from the ``Requested'' state, the processor is restricted to only discussing the identifier of the last document it was asked to process.

If the manager requests that the processor processes document ``doc124'', the conversation returns to the ``Requested'' state, as shown in Figure~\ref{fig:processdocuments-requested2}. The\texttt{?docid} variable has now acquired an updated binding that is now applied to all the transitions using that variable in an immutable context. At this stage, the processor agent may repeat the cycle by processing the document and informing the manager of this, or it may end the conversation by refusing to perform the processing, thus entering the ``End'' state.

\section{Language Integration in Agent Factory} \label{sec:af}

Agent Factory is an extensible, modular and open framework for the development of multi agent systems~\cite{Collier2003}. It supports a number of agent programming languages, including AFAPL/ALPHA~\cite{Collier2001,Collier2005a},
AFAPL2~\cite{Muldoon2009}, AF-TeleoReactive (based on~\cite{Nilsson1994}) and AF-AgentSpeak (an implementation of AgentSpeak(L)~\cite{Bordini2007}). The specific use of ACRE from within AFAPL2 agents has been discussed in a previous paper~\cite{Lillis2010a}.

Agent Factory's implementation of these agent programming languages are based on its \textit{Common Language Framework} (CLF) whereby the way in which sensors, actions and modules are implemented have been standardised across the various languages. This greatly facilitates the integration of additional services into each language, as the core components will be shared. This is the case for ACRE, where minimal effort is required to add support for further languages once one integration has been completed.

The ACRE Architecture consists of a number of components, some of which are platform-independent and some of which require some work to be ported to other platforms and agent programming languages.

\subsection{Protocol Manager}
The \textit{Protocol Manager} (PM) is a platform-independent component that is tasked with making protocols available to agents. When an agent identifies the URL of an ACRE protocol definition it will send this to the PM, which will load the protocol, verify it against the appropriate schema and make it available for interested agents to use. It is also capable of accessing online ACRE repositories that may contain multiple protocol definitions in a centralised location. A repository definition file lists the protocols that it has available. Typically, one PM will exist on an agent platform, so any protocol located by any agent will be shared amongst all agents on the platform (within Agent Factory, the PM is a shared Platform Service). However, there is no technical barrier to individual agents having their own PMs if desired. Previously loaded protocols are also stored locally so that they can be recovered in the event of a platform failure or restart.

\subsection{Conversation Manager}
Whereas the PM is shared amongst agents, each agent has its own \textit{Conversation Manager} (CM), which is used to keep track of the conversations the agent is involved in. The CM monitors both incoming and outgoing communication and matches each message to an appropriate conversation, following the algorithm outlined in Section~\ref{sec:model}. By monitoring the CM, an agent can gain data that can be used to reason about ongoing conversations and the messages it sends and receives. The CM is also platform-independent.

\subsection{Agent/ACRE Interface}
The \textit{Agent/ACRE Interface} (AAI) is specific to the platform and agent programming language being used. This is designed to facilitate the interaction between the agents and the ACRE components mentioned above. The AAI has two distinct principal roles:
\begin{itemize}
   \item To enable an agent to inform the PM and CM of information it holds.
   \item To provide the agent with information about the status and activity of the CM and PM.
\end{itemize}

In the former case, an agent must be capable of informing the PM of the location of any protocols that it wishes to use. This information may originally come from another agent with which it wishes to communicate. The CM requires access to the inbox and outbox of the agent also, so the AAI must provide this service also.

The key role of the AAI is making information about its own communication available to agents. Within Agent Factory, this is done in two complementary ways: \textit{knowledge sensors} and \textit{event sensors}.

A \textit{knowledge sensor} is a sensor that runs on each interpreter cycle of the agent, and provides information on the current state of conversations and protocols. This information currently consists of:
\begin{itemize}
   \item What protocols are already loaded (PM).
   \item For each conversation in which the agent is a participant:
   \item The protocol each conversation is following (CM)
   \item The identity of the other participating agent in each conversation (CM)
   \item The current state of each conversation (CM)
   \item The current status of each conversation (CM)
\end{itemize}

In addition to these, \textit{event sensors} inform the agent whenever events are raised by the PM or the CM. Events currently handled include:
\begin{itemize}
   \item A new protocol has been loaded (PM)
   \item A new conversation has begun (CM)
   \item A conversation has advanced (CM)
   \item A conversation has ended (CM)
   \item An error has occurred in a conversation (CM)
\end{itemize}

In addition to the basic role of information passing, an AAI may augment the capabilities of a language by leveraging the data available from the CM or the PM. For instance, the AAI built for the Agent Factory CLF provides an action of the form \texttt{advance(conversation-id,performative,content)} whereby an agent can advance a specific conversation while providing minimal information. The details about the other participating agent (including its address) are taken from the CM, along with the protocol identifier and content language. Further features in this vein are left for future work.

\section{Conclusions and Future Work} \label{sec:conclusions}

This paper presents an outline of the ACRE conversation reasoning system, how it models protocols and conversations, and how it is integrated into the Agent Factory multi agent framework. Although currently only used with Agent Factory, it is intended that ACRE will be used in conjunction with several other agent programming frameworks and the languages they support. ACRE has been designed to be as language-independent and platform-independent as possible. Despite this, it will be necessary to adapt the system to frameworks and languages other than Agent Factory and the agent programming languages it supports.

Aside from the integration of ACRE into other platforms, future focus will be on the reasoning about conversations at the agent deliberative level and the information that ACRE will need to provide in order to facilitate this. One are of focus will be to allow the grouping of conversations into related groups. As ACRE protocols are limited to two participants, it is necessary to allow agents to relate conversations to each other. One such example will be in a situation where an agent is conducting an auction and has issued a call for proposals to multiple agents. At present, each of these initiates a separate conversation that must be managed by the agent using its own existing capabilities. However, grouping conversations at the Conversation Manager level would allow events to be raised to inform the agent that all conversations in the group had reached the state where a proposal had been received, or left the state where a proposal had been solicited. The key point is the provision of sufficient information for agents to use their deliberative reasoning capabilities in conjunction with the information emanating from the Conversation and Protocol Managers.

In addition, it is intended to explore the ways in which having access to an ACRE layer will add to the native messaging capabilities of other programming languages. This includes the ability to directly advance a conversation, as alluded to in Section~\ref{sec:af}, as well as specifically initiating conversations (rather than relying on message matching on the part of the Conversation Manager). The possibilities are likely to vary with different agent programming languages and it is intended to add these features as appropriate.

The availability of cross-platform communication tools such as ACRE can only aid interoperability between distinct agent platforms, toolkits and programming languages.

\bibliography{LADS2010}

\end{document}